# Second-harmonic generation enhancement in monolayer transition-metal dichalcogenides by an epsilon-near-zero substrate

*Pilar G. Vianna, Aline dos S. Almeida, Rodrigo M. Gerosa, Dario A. Bahamon, Christiano J. S. de Matos\**

*MackGraphe – Graphene and Nanomaterials Research Institute, Mackenzie Presbyterian University, São Paulo - 01302-907, Brazil.*
*\*cjsdematos@mackenzie.br*

Monolayer transition-metal dichalcogenides (TMDCs) present high second-order optical nonlinearity, which is extremely desirable for, e.g., frequency conversion in nonlinear photonic devices. On the other hand, the atomic thickness of 2D materials naturally leads to low frequency converted intensities, highlighting the importance to design structures that enhance the nonlinear response for practical applications. A number of methods to increase the pump electric field at the 2D material has been reported, relying on complex plasmonic and/or metasurface structures. Here, we take advantage of the fact that unstructured substrates with a low refractive index naturally maximize the pump field at a dielectric interface, offering a simple means to promote enhanced nonlinear optical effects. In particular, we measured second harmonic generation (SHG) in $MoS_2$ and $WS_2$ on fluorine tin oxide (FTO), which presents an epsilon-near zero point near our 1550-nm pump wavelength. Polarized SHG measurements reveal an SHG intensity that is one order of magnitude higher on FTO than on a glass substrate.

## Introduction

Much recent attention has been paid to the nonlinear optical response in 2D materials, which are anticipated to provide innovative approaches to enable devices such as all-optical modulators, saturable absorbers, THz wave generators, wavelength converters, optical polarizers, optical sensors and optical limiters[1,2]. These devices suggest applications with new functionalities, high performance, and reduced integration complexity for photonic and optoelectronic platforms[3,4].

2D materials have shown to exhibit extremely high nonlinear optical susceptibilities, with several nonlinear optical effects observed, including second[5–10] and third[9,11,12] harmonic generation (SHG/THG), sum-frequency generation[13], four-wave mixing[13–15], and high- harmonic generation[16–18]. This makes frequency conversion in 2D materials extremely promising, with potential applications to ultrafast pulse characterization[19], all-optical wavelength conversion for telecommunications[4], optical imaging[7,13], quantum information processing[20,21], nanoscale lightsources[22,23], among others. In addition, SHG and THG have proven to be a key technique for crystal orientation and characterization of fundamental material properties[12,24].

In particular, transition-metal dichalcogenide (TMDC) monolayers, such as molybdenum disulfide ($MoS_2$) and tungsten disulfide ($WS_2$) are non-centrosymmetric and present a high second order nonlinear optical susceptibility[5,9], essential for frequency conversion applications based on second-order nonlinear effects. However, direct TMDCs utilization for nonlinear optical applications is still an ongoing challenge due to the atomic thickness of 2D materials and, thus, reduced light-matter interaction, which naturally leads to low net frequency converted intensities. Thus, ways to enhance the process and maximize the nonlinear interaction are crucial for making practical applications viable.

Several strategies have been designed to enhance the nonlinear optical SHG process in TMDCs and other 2D materials, such as excitation near excitonic resonances[25–28], including experiments carried out at low temperature[29] and with electrostatic doping[30]. Another promising class of methods is the combination of the 2D materials with different field-enhancement platforms, including plasmonic nanostructures for localized surface plasmon excitation[31–35], hybrid dielectric structures[36,37], metallic and dielectric metasurfaces governed by bound states in the continuum[38–42], photonic crystal nanocavities[43,44], optical microcavities[45,46] and waveguides[47].

Although these methods have proven to enhance the SHG fields, complex and time-consuming fabrication processes are required. Plasmonic metal nanostructures, for example, require that the fundamental or the frequency-converted field be overlapped with the resonance spectrum of the nanostructures, thus demanding specific nanofabrication techniques to appropriately tune the plasmon resonance[36]. Also, it is well known that noble metal structures exhibit strong optical loss in the visible band, which greatly influences the nonlinear response[35,48]. Metasurfaces, in turn, demand high resolution and often require fabrication by electron-beam lithography, imposing scalability and cost drawbacks to practical manufacturing of the nonlinear devices. The same applies to photonic crystal nanocavities, which rely on high definition lithographic methods, essential for the design of superior quality resonance structures in which mode coupling is required[43,44]. In the case of optical microcavities and waveguides, precise coupling, phase matching and dispersion management are required, thus, adding complexity to practical applications.

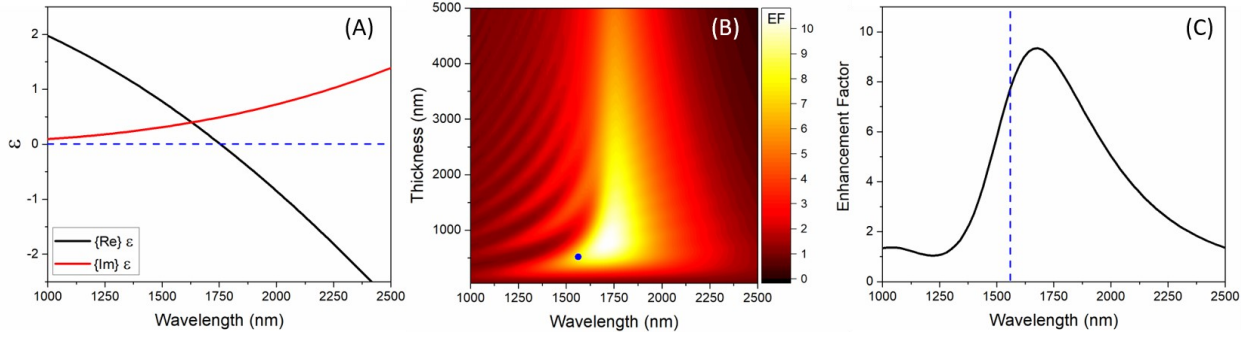

**Fig 1** (A) Dielectric function of FTO. (B) Enhancement factor as function of the wavelength for varying thicknesses of the FTO-on-glass substrate. Blue dot indicates the experimental conditions. (C) Theoretical SHG enhancement factor as a function of the wavelength for a 520-nm FTO thickness. Dashed line shows the experimental wavelength.

A simpler approach to field-enhancement, however, seems to have been so far virtually overlooked. It consists of acknowledging that, as the 2D material lies on the interface between two dielectrics, the pump field on it is given by the electromagnetic boundary conditions that arise from the reflection and refraction phenomena. In particular, for a regular material (i.e., not a metamaterial) substrate, the field will be maximum if the refractive index tends to zero. The impact of a low refractive index, at the pump frequency, can be appreciated from the expression for the second-harmonic intensity, which can be derived using the nonlinear optical sheet susceptibility formalism and that at normal incidence is given by[49]:

$$I_{SHG}(2\omega) = \frac{8 Re\{n(2\omega)\} [2\omega]^2 \left|\chi_S^{(2)}\right|^2 I_1^2(\omega)}{\epsilon_0 c^3 |[1+n(2\omega)][1+n(\omega)]^2|^2}, \quad (1)$$

where, $n(\omega)$ is the substrate refractive index, $\omega$ is the pump frequency and $\chi\_s^{(2)}$ is the 2D material second order sheet susceptibility (see ESI). It is possible to note the quartic dependence on the substrate's refractive index. Clearly, an epsilon-near-zero (ENZ) substrate[50–53] would, thus, maximize SHG.

Here, to increase the nonlinear frequency conversion efficiency in mechanically exfoliated monolayer TMDCs, we propose and demonstrate the use of substrates presenting an epsilon-near-zero point close to the pump wavelength. Fluorine tin oxide (FTO) is used as the substrate with the ENZ point close to the 1550 nm telecommunications spectral range, and $MoS_2$ and $WS_2$ are used as the monolayer 2D materials. Polarized SHG measurements were performed and we observed an SHG intensity 7.6 ± 2.1 times greater for $MoS_2$ and 8.2 ± 2.8 times greater for $WS_2$ on the FTO substrate than that for the same materials on glass, which was found to be compatible with our theoretical predictions.

## Results and Discussion

**Theoretical modelling and SHG enhancement factor**

Theoretical modeling of SHG was planned to reflect the experimental conditions, described later in this letter, in which a monolayer TMDC is deposited onto an FTO film supported by glass. Fig. 1(A) shows the dielectric function of FTO, modeled by the Drude free electron model with parameters obtained by fitting experimental transmission spectra (see Methods and ESI). The ENZ point is expected to be around ~1760 nm. Using the obtained dielectric function, we first calculate the SHG fields generated by $MoS_2$ on the FTO substrate using the Green's function formalism[54] in conjunction with the vacuum interface model[55,56] and the transfer matrix method. We use a variable FTO thickness, as well as the dielectric function of $SiO_2$ available in Ref. [57] for the supporting soda lime glass. The SHG fields above and below the monolayer $MoS_2$ are calculated and then propagated through the different layers using the transfer matrix method (see Methods and ESI). In this way, the total transmitted SHG field was obtained.

For a direct comparison of the SHG on a standard substrate, we normalized the intensity of the SHG on FTO by the intensity of the SHG on glass (see ESI), which we define as the SHG enhancement factor (EF):

$$EF = \frac{I(2\omega)_{FTO}}{I(2\omega)_{glass}} \quad (2)$$

with $I(2\omega)_{FTO}$ being the converted SHG intensity of the $MoS_2$ monolayer deposited on FTO supported by glass and $I(2\omega)_{glass}$ the SHG intensity of $MoS_2$ on glass. It is also important to mention, that, as the complex dielectric function of the substrate is used in our model for Fresnel coefficient calculations, optical absorption is fully accounted for (see ESI).

Fig. 1B shows the EF as a function of the wavelength for different FTO thicknesses. It is possible to observe that the SHG intensity varies with the substrate thickness, due to interferometric effects, and excitation wavelength, achieving a maximum EF of 10.68 at 1720 nm wavelength for a 770-nm-thick substrate. Fig. 1C shows the EF as a function of the pump wavelength for our experimental substrate (520 nm thickness), represented by the blue circle in Fig. 1(B). For our pump, at 1560 nm, represented by the blue dashed line in Fig. 1C, the SHG on FTO is ~7.7 times more intense than that on glass. The largest enhancement factor, of ~9.3, is observed at 1680 nm, for the 520-nm thickness.

**Experiments**

Experimentally, monolayer $MoS_2$ and $WS_2$ were obtained by mechanical exfoliation and transferred to the surfaces of FTO, with 520 nm thickness on glass, and plain glass (for comparison) (for additional details, see Methods). Flake characterization was performed by Raman spectroscopy (WITec Alpha 300R) with excitation at 532 nm (see ESI). To minimize possible substrate thickness/roughness variations, $MoS_2$ and $WS_2$ flakes were transferred to the same FTO substrate (see Fig. S7 in ESI). The glass used as the substrate in the reference experiments was the same as the one used for FTO deposition.

Nonlinear optical measurements consisted of SHG. The experimental setup can be seen in Fig. 2 and included a 1560 nm mode-locked Er-doped fiber laser (FFS, Toptica Photonics) with 89 MHz repetition rate and 150 fs pulse duration, used as the pump. The laser was focused on the sample at normal incidence with a 20× objective lens, leading to a ~3 μm focused beam diameter. Using a pair of 5× objective lenses, the SHG signal at 785nm was collected by transmission and directed to a spectrometer equipped with an iDus Si CCD detector. The detector presents no quantum efficiency at 1560 nm and the remnant pump power was low after the sample to avoid any possible damage to the spectrometer. The spectrometer's grating and settings, set to measure the SHG and THG wavelengths, did not direct the pump radiation to the detector. Therefore, the pump generated no features in the spectrum and no filter was needed for blocking the laser before signal collection. Polarized SHG measurements were performed using a linear polarizer (for the pump) and an analyzer (for the SHG signal) in motorized rotation stages. A quarterwave plate adjusted the pump polarization to be circular, so that rotating the polarizer led to minimal intensity changes. A light source, a CCD camera and some optics were used to optically image, in reflection, the monolayer flakes and the THG signal (at 520 nm), with the latter indicating the pump beam position (see ESI for additional details). The pump average power was kept at 10 mW to avoid damage of the monolayer TMDs. Measurements were performed with the analyzer parallel and perpendicular to the incident light polarization.

For polarization-resolved measurements, the two motorized stages, containing the polarizer and the analyzer, were rotated with 5º steps, yielding 73 SHG spectra measured from 0º to 360º. Data for $MoS_2$ on FTO and $MoS_2$ on glass were acquired at the same day to minimize system induced intensity fluctuations, and the standard deviation of at least 10 measurements was used for error estimates. Spectra were obtained with 2 s integration time and 2 accumulations. Each spectrum was individually analyzed and the SHG intensity values calculated from the integral of each spectrum.

The SHG intensity dependence on the polarization angle can be seen in Fig. 3, for the parallel pump-SHG polarization configuration. The curves for bare FTO, $MoS_2$/FTO and $WS_2$/FTO samples are shown in black, red and blue, respectively. All data were normalized by the $MoS_2$/Glass maximum SHG intensity. The maximum normalized SHG intensities were found to be 29.3, 13.3 and 5.6 for $WS_2$/FTO, $MoS_2$/FTO and FTO respectively. It is important to note that FTO presented a relatively strong SHG component, which can only be observed in the parallel polarization configuration. This FTO SHG component is possibly an evidence of some substrate local remnant (and variable) crystallinity[58], even though the film was produced by sputtering (see Methods), which tends to yield an amorphous material. This component is not uniform along the substrate and interferes with the TMDCs SHG signal, originating the distorted SHG patterns in the parallel configuration shown in Fig. 3, and making it difficult to separate the substrate and the 2D material contributions (see ESI). Without the substrate's interference, the SHG signal in monolayer TMDCs is expected to exhibit the typical sinusoidal dependence on the polarization angle, with a 60º period, compatible with the hexagonal symmetry and point group $D_{3h}$[4,10]. It is important to emphasize that FTO SHG component is only observed in the parallel configuration, and for that reason, our analysis mainly focuses on the SHG obtained with the perpendicular pump-SHG polarization configuration, which eliminates the FTO interfering signal.

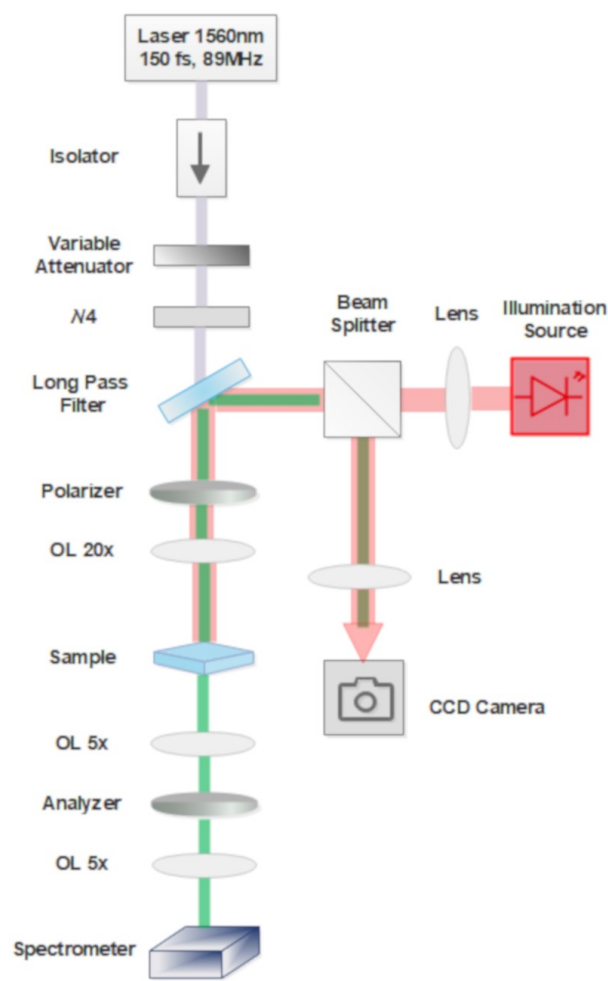

**Fig. 2** Experimental setup for SHG characterization of the samples.

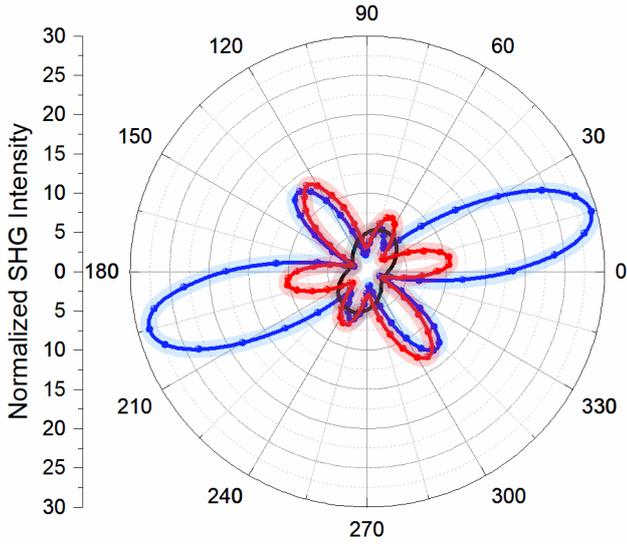

**Fig. 3:** Experimental SHG intensity as a function of pump polarization angle in the parallel polarization configuration for FTO (black), MoS$_2$/FTO (red) and WS$_2$/FTO (blue), normalized by the maximum measured SHG signal on the MoS$_2$/glass sample.

In Fig. 4, the polarized SHG measurements for the perpendicular pump-SHG polarization configuration can be observed for both the TMDCs on FTO and on glass. The intensity values were again normalized by the maximum SHG intensity obtained with MoS$_2$ on glass. Fig. 4a presents the experimental results for MoS$_2$, with the red (black) curve representing results for the FTO (glass) substrate. The SHG intensity is 7.6 ± 2.1 times larger for the material deposited on FTO than that for the monolayer on glass, agreeing with theoretical predictions. The polarization-dependent SHG profile for MoS$_2$ in the perpendicular configuration presents the characteristic sinusoidal behaviour with a 60º period, as expected for monolayer MoS$_2$ (see ESI for theoretical prediction). MoS$_2$/Glass and WS$_2$/Glass in the parallel configuration present the same pattern and intensity as their counterparts in the perpendicular configuration, only with a 30° phase shift (see ESI).

The same measurements were performed for WS$_2$ deposited on FTO and glass, shown as the blue and black plots in Fig. 4b, respectively. The results indicate an SHG enhancement of 8.2 ± 2.8 times comparing FTO and glass as substrates. In the case of WS$_2$, strain is believed to have yielded the amplitude asymmetry observed in the polarization-resolved SHG intensity plots[59–61]. The origin of strain is usually attributed to the inherent lack of stiffness of PDMS used on sample preparation. PDMS being soft can get slightly deformed during transfer by the pressure exerted upon contact with the target substrate, likely being the deformation source to induce strain in the flake being transferred[60,62]. In principle, and as already observed by others[63–66], the photoluminescence (PL) peak shift can be considered an indicator of how the electronic band structure is altered by the application of strain in TMDCs. In our case, we observed a shift of 0.1 eV on the WS$_2$ PL peak comparing the flakes before and after the transfer process (see ESI), compatible with strain of less than 1% in monolayer WS$_2$[63].

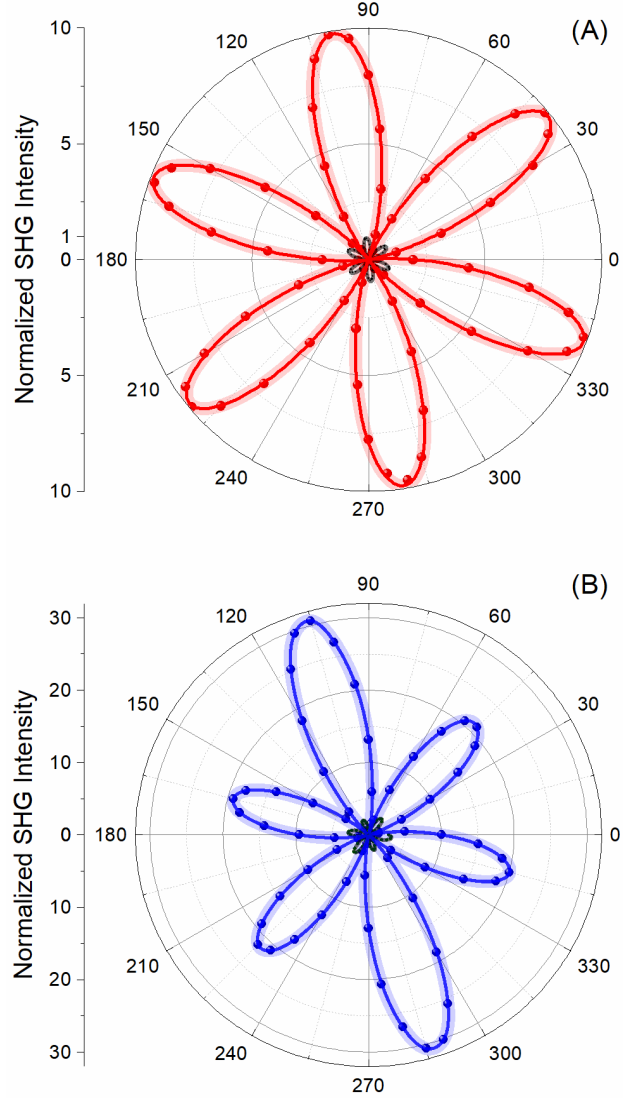

**Fig. 4:** Experimental SHG intensity as a function of pump polarization angle in the perpendicular polarization configuration for (A) MoS$_2$/FTO (red) and MoS$_2$/Glass (black); and for (B) WS$_2$/FTO (blue) and WS$_2$/Glass (black).

Nevertheless, it is possible to observe the same enhancement values, within the errors, for monolayer MoS$_2$ and WS$_2$ on FTO, as indeed expected and confirming that the enhancement mechanism is only dependent on the ENZ substrate, which provides a simple and easily scalable alternative for nonlinear optical wavelength conversion.

## Conclusions

We reported on SHG in mechanically exfoliated monolayer TMDCs enhanced by an epsilon-near-zero (ENZ) substrate. This substrate presents itself as an extremely simple and low-cost alternative to enhance the pump electric field at 2D material, thus increasing the frequency converted intensity. By optimizing the substrate thickness and operating with a pump near the ENZ point, it is possible to enhance the nonlinear optical fields, with increased potential for practical nonlinear applications. The converted signal was experimentally found to be 7.2 ± 2.1 and 8.2 ± 2.8 times higher on the ENZ substrate than on glass, for $MoS_2$ and $WS_2$, respectively, agreeing with theoretical predictions.

## Methods

**Theoretical model.** Extended description of the theoretical model can be found in the ESI. Briefly, the sample is modeled as a multilayer system with the z-direction normal to the surface. Layer 1 is a semi-infinite vacuum region ($z > 0$), where the pump source is located. Layer 2 corresponds to the FTO substrate of thickness $d$ ($0 > z > -d$), with a monolayer $MoS_2$ sheet on top ($at\ z = 0$); and layer 3 is a semi-infinite glass slide with $z < -d$, where the detector is placed. To evaluate the SHG fields in layer 3, the pump fields are first calculated. Then, the SHG fields above and below the $MoS_2$ are calculated and subsequently propagated through the different layers. In this way, the transmitted SHG field in layer 3 was obtained. Green's function formalism, with the infinitesimal vacuum gap approach, is used to obtain the SHG fields radiated by the monolayer $MoS_2$ on FTO and on glass. To compare the field intensity on both substrates, the $MoS_2$/FTO SHG intensity was normalized by the $MoS_2$/glass SHG intensity (see ESI for detailed models). The dielectric function of FTO was described by the Drude free electron model, where $\varepsilon_\infty$ = 2.95, $\omega_p$ = 1.89×10$^{15}$ s$^{-1}$ and $\gamma$ = 0.9×10$^{14}$ s$^{-1}$, which were obtained by fitting experimental transmission curves (see ESI). The glass refractive index was taken to be given by[57] $n = 1.5130 - 0.003169\lambda^2 + \frac{0.003962}{\lambda^2}$.

**Sample preparation and characterization.** $MoS_2$ and $WS_2$ monolayer flakes were mechanically exfoliated onto UV-ozone treated polydimethylsiloxane (PDMS) and transferred to the surfaces of FTO and glass (for comparison). The sputter coated FTO substrate was kindly provided by MSE Supplies LLC (TEC™ 7, 7-8 Ohm/sq., on a 2.2-mm thick soda lime glass) with a root mean square (RMS) roughness of approximately 17.8 nm, as determined by atomic force microscopy (AFM). The substrate was polished, which reduced the intrinsic roughness to 1.7 nm, allowing for a better adhesion of the monolayer flakes during the transfer process (see Fig. S4 in ESI). Substrate thickness before and after polishing was determined by interferometric measurements (3D laser scanning microscope - model VK-X200 at 408 nm), yielding 610 nm and 520 nm, respectively.

Flake characterization was performed by Raman spectroscopy (WITec Alpha 300R) for both $MoS_2$ and $WS_2$ with 532 nm excitation (see Figs. S5 and S6 in ESI). After the flakes were transferred to the substrates (FTO and glass), samples were annealed in vacuum at 200ºC for 2h followed by additional vacuum annealing for 3h at 200ºC to minimize polymer residue and strain[60]. To minimize possible substrate thickness/roughness variations, $MoS_2$ and $WS_2$ flakes were transferred to the same FTO substrate. The glass sample was the same as the one used for FTO deposition.

## Conflicts of interest

There are no conflicts to declare.

## Acknowledgements

This work was funded by FAPESP (Thematic Projects 2015/11779-4 and 2018/25339-4), the Brazilian Nanocarbon Institute of Science and Technology (INCT/Nanocarbon – 88887.195743/2018-00), CNPq, FAPEMIG and CAPES-PRINT (Grant 88887.310281/2018-00). P.G.V was supported by CAPES scholarships (grant nos. 88887.195743/2018-00 and 88887.370132/2019-00). The FTO substrates were kindly provided by MSE Supplies LLC (https://www.msesupplies.com). The use of the 3D laser scanning microscope - model VK-X200 at LNNano facilities for FTO thickness measurement (process: DSF-LD-25815) is gratefully acknowledged.

# Supporting Information for:
# "Second-harmonic generation enhancement in monolayer transition-metal dichalcogenides by an epsilon-near-zero substrate"

Pilar G. Vianna, Aline dos S. Almeida, Rodrigo M. Gerosa, Dario A. Bahamon, Christiano J. S. de Matos*

MackGraphe – Graphene and Nanomaterials Research Institute, Mackenzie Presbyterian University, São Paulo - 01302-907, Brazil.

* E-mail: cjsdematos@mackenzie.br

# 1. Second-harmonic generation intensity

To quantify the second-order nonlinear response of a 2D material layer, a monolayer of the material can be considered a nonlinear polarization sheet ($\chi_s^{(2)}$) at the interface between two media, in our case, air and the FTO substrate. The second harmonic generation process will depend not only on the nonlinear polarization, but also on the dielectric properties of the surrounding media, which impact the amplitude of the electric field at the interface, as determined by the electromagnetic boundary conditions. With the pump at normal incidence and considering a frequency-dependent refractive indices, the second-harmonic intensity is given by[1]:

$$I_{SHG}(2\omega) = 2\epsilon_0 Re\{n_2(2\omega)\}c \left| \frac{4i(2\omega)\, n_1^2(\omega)\, \overleftrightarrow{\chi_s}^{(2)} \hat{e}\hat{e}\, E_1^2(\omega)}{c[n_1(\omega) + n_2(\omega)]^2[n_1(2\omega) + n_2(2\omega)]} \right|^2, \quad (1)$$

where $\omega$ is the pump frequency, $n_1$ is the refractive index of the incidence medium, and $n_2$ is the substrate refractive index, $\overleftrightarrow{\chi_s}^{(2)}$ is the second-order nonlinear sheet susceptibility tensor, $\hat{e}$ is the unit vector associated to the polarization of the pump field, $c$ is the speed of light in vacuum and $\epsilon_0$ is the vacuum permittivity. Note that the refractive indices are taken to be complex. As the medium of incidence is air, we make $n_1(\omega) = n_1(2\omega) = 1$. In addition, with $|E_1(\omega)|^2 = \frac{I_1(\omega)}{2\,\epsilon_0\, n_1(\omega)\, c}$, and redefining $n_2$ as $n$ to avoid confusion with the nonlinear refractive index, we get:

$$I_{SHG}(2\omega) = \frac{8Re\{n(2\omega)\}\,[2\omega]^2 \left|\overleftrightarrow{\chi_s}^{(2)}\hat{e}\hat{e}\right|^2 I_1^2(\omega)}{\epsilon_0 c^3\, |[1 + n(2\omega)][1 + n(\omega)]^2|^2}. \quad (2)$$

with $I_1(\omega)$ being the pump intensity at the incidence, and in which $\left|\overleftrightarrow{\chi_s}^{(2)}\hat{e}\hat{e}\right|^2$ can be replaced with $\left|\chi_s^{(2)}\right|^2$, with $\chi_s^{(2)}$ representing the relevant tensor components.

# 2. Theoretical model

The dielectric function of FTO was described by the Drude free electron model:

$$\varepsilon = \varepsilon_\infty - \frac{\omega_{p2}}{\omega_2 + i\gamma\omega} \quad (3)$$

where $\varepsilon_\infty = 2.95$, $\omega_p = 1.89\times10^{15}$ s$^{-1}$ and $\gamma = 0.9\times10^{14}$ s$^{-1}$ were obtained by fitting experimental transmission spectra. These fittings are shown in Figures 1(A) and 1(B) for the *p* polarization at 0° and 45°, respectively, while Figures 2(A) and 2(B) show the fittings for the *s* polarization at 0° and 45°, respectively.

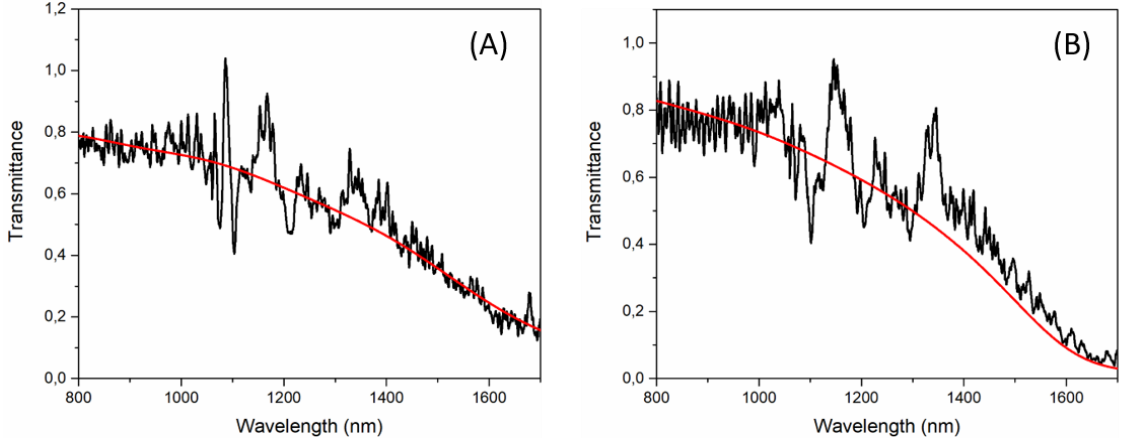

**Figure S1:** Experimental (black) and theoretical (red) transmittance for the *p* polarization at 0° (A) and 45° (B).

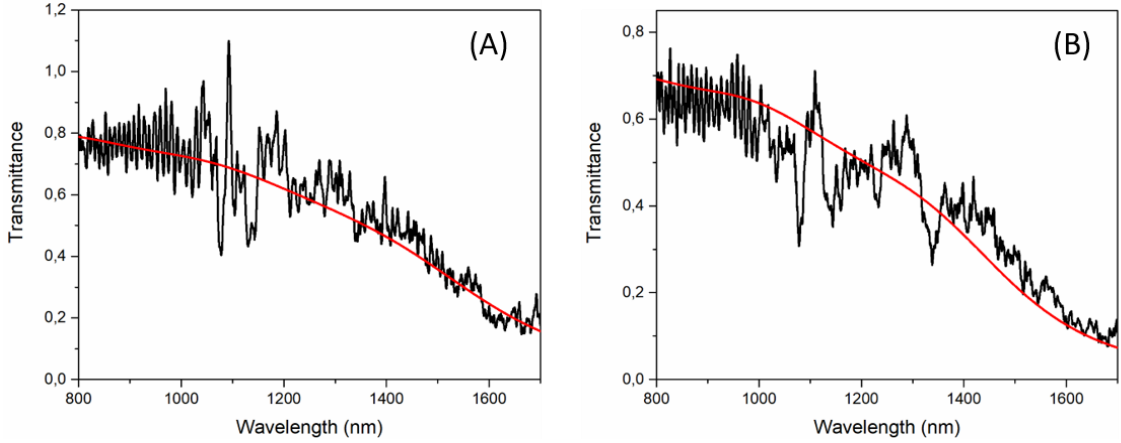

**Figure S2:** Experimental (black) and theoretical (red) transmittance for the *s* polarization at 0° (A) and 45° (B).

The sample is modeled as a multilayer system with the z-direction normal to the surface. Layer 1 is a semi-infinite vacuum region ($z > 0$) where the pump source is located. Layer 2 corresponds to the FTO substrate of thickness $d$ ($0 > z > -d$), with a monolayer MoS$_2$ sheet on top (at $z = 0$) and layer 3 is a semi-infinite glass slide with $z < -d$, where the detector is placed. To evaluate the SHG fields in layer 3, the fundamental fields are first calculated. At normal incidence, the fundamental field at the MoS$_2$ sheet $\vec{E}^\omega = E^\omega \hat{x} = (1 + r_{13}^\omega)E_i^\omega \hat{x}$ is related to the incident pump field ($E_i^\omega$) by the Fresnel reflection coefficient:

$$r_{13}^\omega = r_{12} + \frac{t_{12}t_{21}r_{23}e^{i2k_2d}}{1 - r_{21}r_{23}e^{i2k_2d}}. \qquad (4)$$

Note that the coefficients are evaluated at ω, with $r_{12} = \frac{n_1 - n_2 - Z_0\sigma^s}{n_1 + n_2 + Z_0\sigma^s}$ and $t_{12} = \frac{2n_1}{n_1 + n_2 + Z_0\sigma^s}$, which accounts for the monolayer MoS$_2$ contribution through the sheet response[2] $Z_0\sigma^s = \left(\frac{1}{\epsilon_0 c}\right)[-i(\epsilon - 1)\epsilon_0\omega d]^2$, where $d = 0.65$ Å, is the effective thickness of the monolayer MoS$_2$ sheet, $\epsilon = 4.97$ is the dielectric function, $\epsilon_0$ is the vacuum permittivity, and $n_1$, $n_2$ and $n_3$ are the refractive indexes for air, FTO and glass, respectively.

Next, the Green's function formalism[3] with the infinitesimal vacuum gap approach[4,5] is used to obtain the SHG fields radiated by the monolayer MoS$_2$. Using this approach, an infinitesimal vacuum gap is introduced between the MoS$_2$ layer (at $z = 0^+$) and the FTO layer ($0^- > z > -d$). The suspended MoS$_2$ sheet generates a downward (-) and an upward (+) SHG field[3], defined as:

$$\vec{E}_\pm^\Omega = i\frac{K_0}{2\epsilon_0}\left(P_x^\Omega \hat{x} \mp P_y^\Omega \hat{y}\right)e^{\pm iK_0|z|}, \qquad (5)$$

where $K_0 = \frac{\Omega}{c} = \frac{2\omega}{c} = 2k_0$, $P_x^\Omega = \epsilon_0 \chi_s^{(2)} \cos(3\theta)\left((1 + r_{13}^\omega)E_i^\omega\right)^2$ and $P_y^\Omega = \epsilon_0 \chi_s^{(2)} \sin(3\theta)\left((1 + r_{13}^\omega)E_i^\omega\right)^2$. Here, $\theta$ is the angle between the crystallographic $x'$ direction (armchair) and the laboratory $x$ coordinate; $\chi_s^{(2)} = \chi_{s\ y'y'y'}^{(2)} = -\chi_{s\ y'x'x'}^{(2)} = -\chi_{s\ x'x'y'}^{(2)} = -\chi_{s\ x'y'x'}^{(2)}$ is the second order sheet susceptibility. In this case, we have assumed a time dependence $e^{-i\Omega t}$ and defined the $\hat{s} = \hat{x}$ and $\hat{p}_\pm = \hat{s} \times (\pm\hat{K}_0) = \mp\hat{y}$ polarization. Both polarizations have been treated separately, $u = s, p$, and represent the field in the "m$_{th}$" layer by a two-element column vector $e_{m,u}$, where the top (bottom) element described the upward (downward) propagation directions. Therefore, the fields above and below MoS$_2$ can be written[3] as $e_{1u}(0^+) = v_u + e_{1u}(0^-) = v_u + M_{13}^u e_{3u}(-d)$. Here, $v_u$, is an SHG field discontinuity introduced by the monolayer, and $M_{13}^u$ is the transfer matrix. Given that there is no downward (upward) propagation of the SHG field for $z > 0^+$ ($z > d$), for the s polarization we have:

$$\begin{bmatrix} E_{1x}^\Omega(0^+) \\ 0 \end{bmatrix} = i\frac{K_0}{2\epsilon_0}P_x^\Omega \begin{bmatrix} 1 \\ -1 \end{bmatrix} + M_{13}^x \begin{bmatrix} 0 \\ E_{3x}^{\Omega-}(-d) \end{bmatrix} \qquad (6)$$

With the transfer matrix defined as:

S4

$$M_{13}^x = \frac{1}{T_{13}} \begin{bmatrix} T_{13}T_{31} - R_{13}R_{31} & R_{13} \\ -R_{31} & 1 \end{bmatrix} \quad (7)$$

It is important to highlight that:

i. The transfer matrix $M_{13}^u$ is defined between layer 1 at $z = 0^-$ and layer 3, so that the transmission $(T_{ij})$ and reflection $(R_{ij})$ terms do not include the MoS$_2$ sheet response;
ii. The elements of the transfer matrix are evaluated at $\Omega = 2\omega$. Using Equation 6 for the s polarization and a similar equation for the p polarization, the field in layer 3, is obtained:

$$\vec{E}_3^\Omega = i \frac{K_0}{2\epsilon_0} \left( T_{13} P_x^\Omega \hat{x} + T_{13} P_y^\Omega \hat{y} \right) e^{-iK_0|z|} \quad (8)$$

## 2.1 Monolayer MoS$_2$ on glass

To compare the SHG fields of the MoS$_2$/FTO and MoS$_2$/glass configurations, the SHG fields for MoS$_2$ on glass were also calculated. The MoS$_2$/glass sample model has two layers, in which layer 1' is a semi-infinite vacuum layer ($z > 0$) and layer 2' ($z < 0$) is the glass substrate with the monolayer MoS$_2$ on top. Following the previously detailed methods, we obtain the SHG field in the glass region:

$$\vec{E}_{2'}^\Omega = i \frac{K_0}{2\epsilon_0} \left( T_{1'2'} P_x^\Omega \hat{x} + T_{1'2'} P_y^\Omega \hat{y} \right) e^{-iK_0|z|} \quad (9)$$

Where, $T_{1'2'} = \frac{2n_{1'}}{n_{1'}+n_{2'}}$ with $n_{1'}$ and $n_{2'}$ evaluated at $\Omega = 2\omega$, and $P_x^\Omega = \epsilon_0 \chi_s^{(2)} \sin(3\theta) \left( (1 + r_{1'2'}^\omega) E_i^\omega \right)^2$ and $P_y^\Omega = \epsilon_0 \chi_s^{(2)} \sin(3\theta) \left( (1 + r_{1'2'}^\omega) E_i^\omega \right)^2$. As previously reported, $r_{1'2'}^\omega$ is evaluated at the fundamental frequency and include the MoS$_2$ sheet response.

## 2.2 Normalized intensity and optical absorption

In order to compare the transmitted field intensity using both substrates, the MoS$_2$/FTO SHG intensity $(I(2\omega)_{FTO})$ was normalized by the MoS$_2$/glass SHG intensity $(I(2\omega)_{glass})$, which we define as the enhancement factor (EF):

$$EF = \frac{I_{123}}{I_{1'2'}} = \frac{I(2\omega)_{FTO}}{I(2\omega)_{glass}} = \frac{|T_{13}(1+r_{13}^{\omega})^2|^2}{|T_{1'2'}(1+r_{1'2'}^{\omega})^2|^2} \tag{10}$$

Given the large wavelength in the FTO substrate around the ENZ condition, the reflection coefficient $r_{13}^{\omega}$ remains nearly constant as a function of the dielectric thickness of the FTO substrate. Consequently, the EF dependence on the substrate thickness is given by the transmission function $T_{13}$. Our results remain valid for a broad range of thicknesses as can be observed in Figure 1B (Main text).

It is also important to mention that, as the complex dielectric function of the substrate is used in our model for Fresnel coefficient calculations, absorption is fully accounted for. Figure S3 presents the calculated transmission $|t_{13}^{\omega}|^2$, reflection $|r_{13}^{\omega}|^2$ and absorption $A = 1 - |r_{13}^{\omega}|^2 - |t_{13}^{\omega}|^2$ for the fundamental frequency as function of the thickness. Additionally, the classical skin depth $\delta = c/\omega\kappa(\omega)$ for FTO is superimposed as the black curve in each panel.

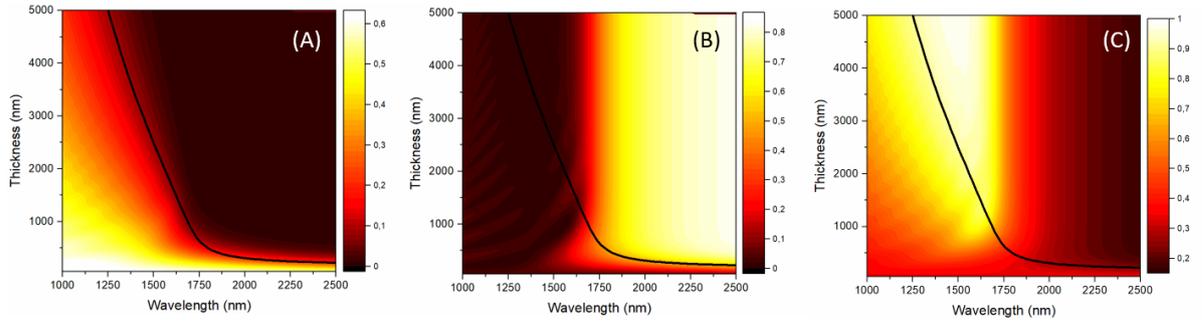

**Figure S3.** (A) Transmission coefficient $|t_{13}^{\omega}|^2$. (B) Reflection coefficient $|r_{13}^{\omega}|^2$. (C) Absorption $A = 1 - |r_{13}^{\omega}|^2 - |t_{13}^{\omega}|^2$. The classical skin depth is represented as the black curve in each panel.

From Figure S3, it is possible to observe that the optical absorption around the ENZ point becomes significant for substrates thicker than 1000 nm. Comparing the optical absorption of the ENZ substrate with the EF map (Figure 1B – Main text), it is evident that an increase in the absorption, reduces the EF. In fact, balance between reflection and moderate absorption around the ENZ point is what defines the observed region with high EF values. For the largest enhancement factor of 9.3 at 1680 nm and a thickness of 520 nm, as in our substrate, we have $|r_{13}^{\omega}|^2 = 0.32$, $A = 1 - |r_{13}^{\omega}|^2 - |t_{13}^{\omega}|^2 = 0.47$ and $\delta = 1046.5$ nm; which is evidence that, despite the substrate absorption around the ENZ point, its thickness is smaller than the classical skin depth and the field is not completely attenuated.

# 3. FTO substrate characterization and processing

FTO (Fluorine doped tin oxide) deposited on SLG (soda lime glass) commercial samples were kindly provided by MSE Supplies LLC. The original FTO thickness was approximately 600 nm with a sheet resistivity of 7-8 Ω/□. From AFM topography measurements, Figure S3(A), the films were found to have a root mean square (RMS) roughness of 17.8 nm, which virtually prevented the transfer of monolayer transition metal dichalcogenide (TMD) flakes. Therefore, the FTO substrates were polished with fine-grain diamond sandpaper (1 µm grit for 3 min, 0.5 µm grit for 5 min and 0.1 µm grit for 5 min). The AFM topography after polishing can be observed in Figure S3(B), corresponding to an RMS roughness of approximately 1.7nm and a measured thickness of 520 nm.

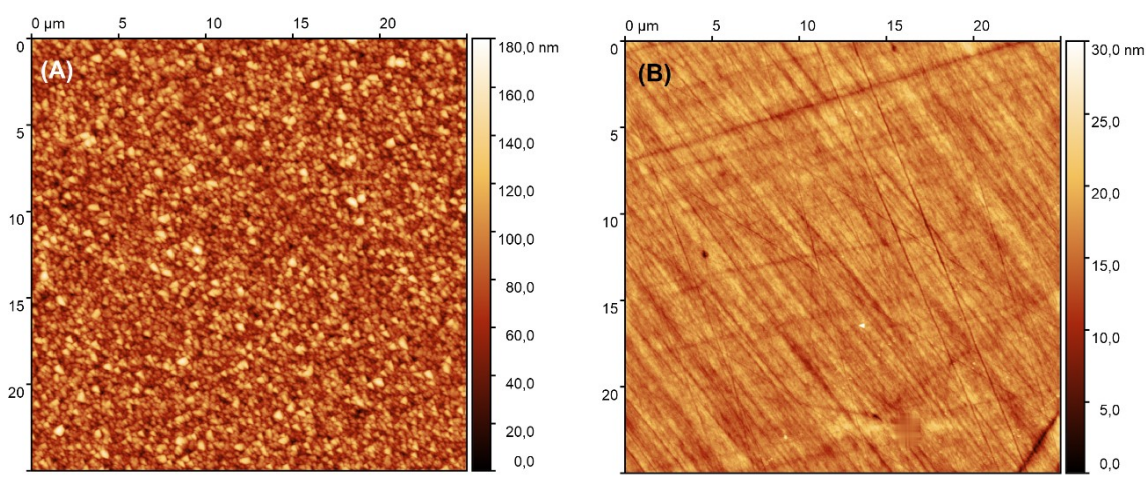

**Figure S4:** AFM topography images for the original FTO substrate (A) and for the polished FTO substrate (B).

# 4. MoS₂ and WS₂ Raman and optical microscopy characterization

Raman spectroscopy characterization was performed to confirm the flake thicknesses. Optical microscopy images of the $MoS_2$ (A) and $WS_2$ (C) flakes on glass can be observed in Figure S4 and $MoS_2$ (A) and $WS_2$ (C) on FTO can be seen in Figure S5.

Figure S4(B) compares spectra from the monolayer and bulk $MoS_2$ regions of the flake in Figure S4(A), with a wavenumber difference between the $A_{1g}$ to $E^1_{2g}$ modes of approximately 18.6 cm⁻¹ and 24.5 cm⁻¹, respectively, which is compatible with the literature for monolayer and bulk $MoS_2$ flakes[6]. The same features were observed for $MoS_2$ deposited on FTO, as shown in Figure S5(B).

For the $WS_2$ flakes, the Raman signature can be observed in Figures S4(D) for the material deposited on the glass substrate and in Figure S5(D) for the 2D material deposited on FTO. It is possible to observe characteristic monolayer feature $I_{[2LA(M)+E_{2g}^1]}/I_{A_{1g}} > 2$ at the 532-nm wavelength resonant excitation condition, with $I_{[2LA(M)+E_{2g}^1]}$ being the sum of the intensities in Raman modes $2LA(M)$ and $E_{2g}^1$ and $I_{A_{1g}}$ being the intensity of Raman mode $A_{1g}$ [7].

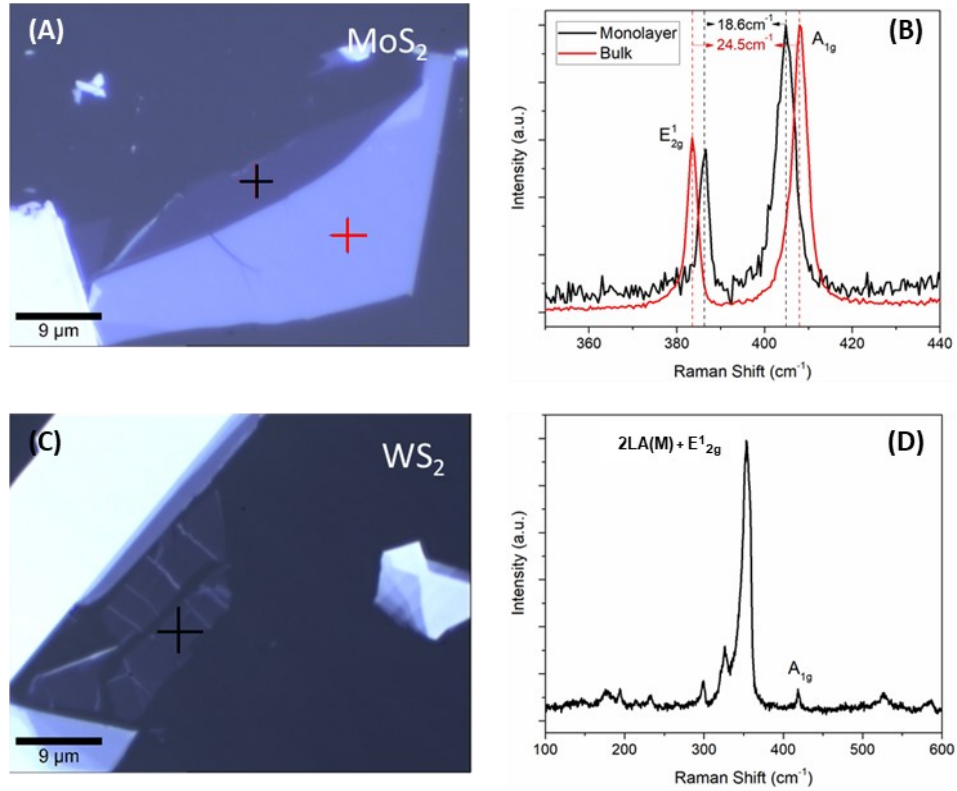

**Figure S5:** MoS$_2$ and WS$_2$ on glass optical characterization. Optical microscope images of MoS$_2$ (A) and WS$_2$ (C) deposited on glass. (B) & (D) Raman spectra obtained at the same color positions marked by a cross in (A) and (C), respectively. Raman data obtained at 532 nm with 0.5 s, 10 accumulations and 3.5 mW laser power for MoS$_2$; and 2 s integration time, 10 accumulations and 1.15 mW laser power for WS$_2$.

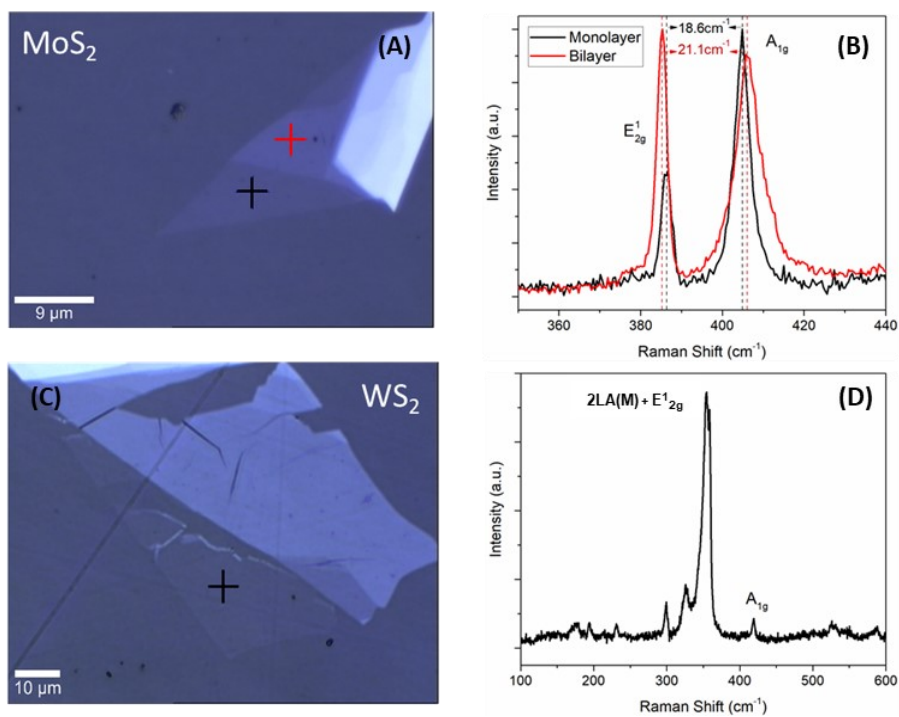

**Figure S6:** MoS$_2$ and WS$_2$ on FTO optical characterization. Optical microscope images of MoS$_2$ (A) and WS$_2$ (C) deposited on FTO. (B) & (D) Raman spectra obtained at the same color positions marked by a cross in (A) and (C), respectively. Raman data obtained at 532 nm with 2 s, 10 accumulations and 3.5mW laser power for MoS$_2$; and 2 s integration time, 10 accumulations and 1.15 mW laser power for WS$_2$.

To minimize possible substrate roughness variations, MoS$_2$ and WS$_2$ flakes were transferred to the same FTO substrate, shown in the optical microscope image of Figure S6. For the samples of TMDs on glass, the glass substrate was the same as the one used for FTO deposition.

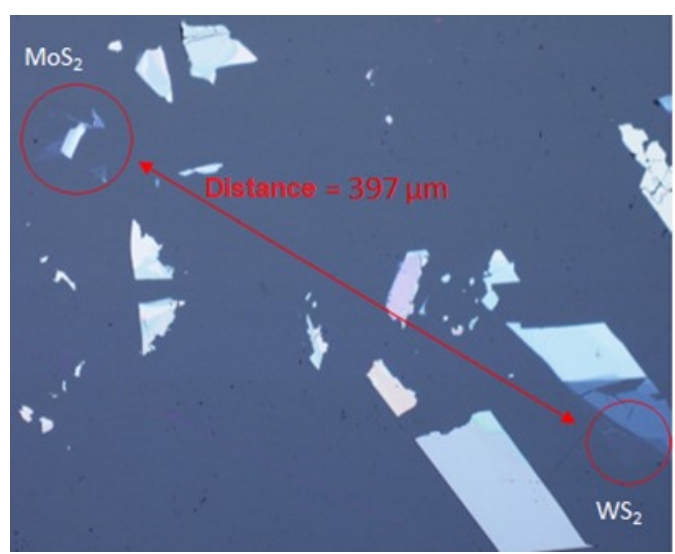

**Figure S7:** Optical microscope image of MoS$_2$ and WS$_2$ flakes deposited on the same FTO substrate. The red circle highlights the monolayer regions.

## 5. Sample imaging and flake location

Reflected THG can be observed overlaid on the linear optical image, shown in Figure S8. A red LED source, a CCD camera and some optics were used to optically image, in reflection, both the TMDC flakes and the THG beam profile generated by the pump laser. A neutral density filter is used to control the red LED light intensity (more attenuation on Figure S8(A) and less attenuation on Figure S8(B)), which allows for a better visualization of the THG spot. Also, adjusting the CCD camera's RGB response improves visualization. THG arises from both FTO and the 2D material (with the spot getting brighter on flakes), which is not a problem, since it is simply used for visualization. The yellow triangle in Figure S8(A) delimits the monolayer flake, which can be more easily seen in Figure S8(B).

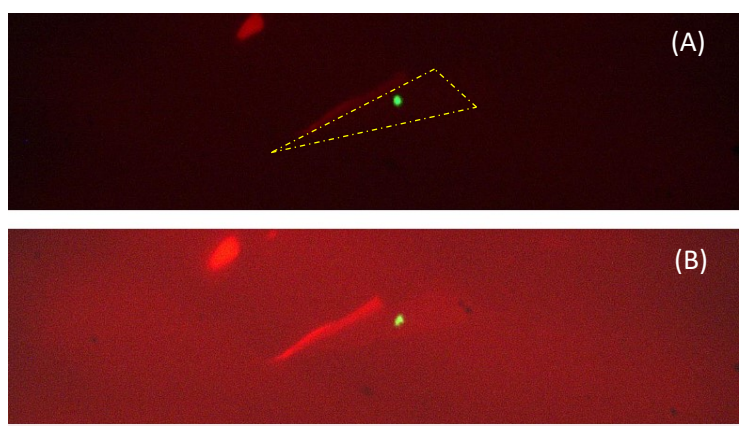

**Figure S8.** Optical image of the flakes under red LED light illumination and the THG beam. (A) Attenuated red LED light. The yellow triangle delimits the monolayer TMDC flake. (B) Higher LED light intensity on the sample.

## 6. SHG polarization dependence in FTO

As the FTO is deposited by sputtering, it tends to be amorphous, with the observed SHG component possibly arising from some local remnant (and variable) crystallinity[8,9]. As a consequence, the SHG polarization dependence (parallel configuration) is variable (remaining inexistent in the perpendicular configuration), as shown in Figure S9, for multiple positions of the same substrate. For that reason, as mentioned in the manuscript, the analysis mainly focuses on the perpendicular polarization configuration, considered interference-free from the FTO SHG.

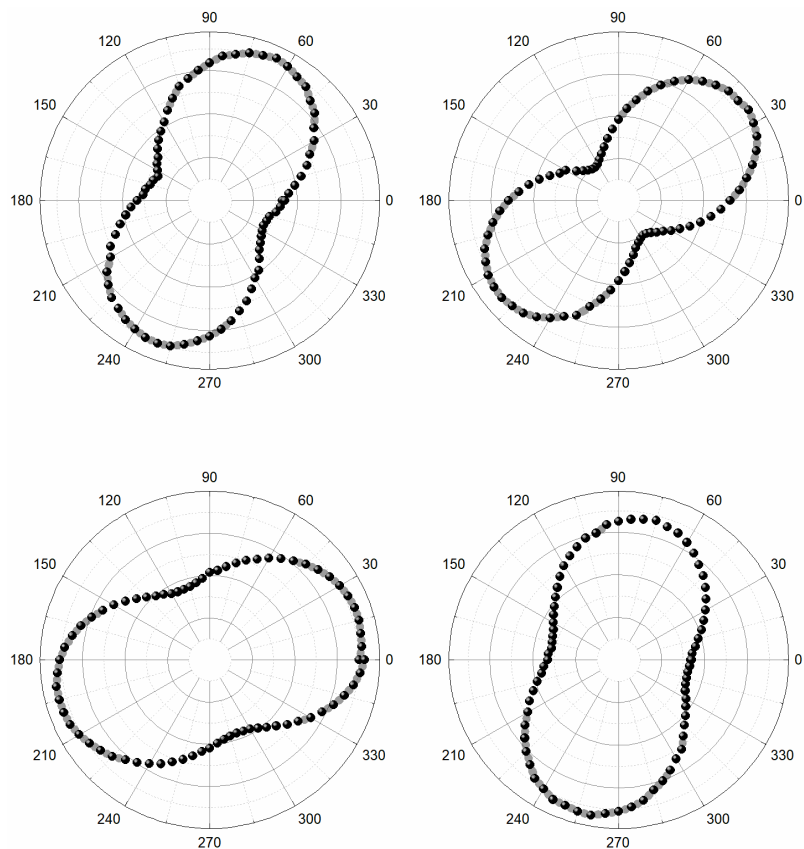

**Figure S9.** SHG intensity as a function of pump polarization angle in the parallel polarization configuration for FTO at different positions of the same substrate.

## 7. Theoretical curves for the SHG polarization dependence in the perpendicular configuration

For MoS$_2$ in the perpendicular configuration, the theoretical prediction is shown below, in Fig.S10. The red plot represents the SHG intensity as a function of the pump polarization angle for MoS$_2$/FTO and the black plot for MoS$_2$/Glass. Data has been shifted by 9° and 6° in the FTO and Glass plots, respectively, to reflect the experimental crystal orientations.

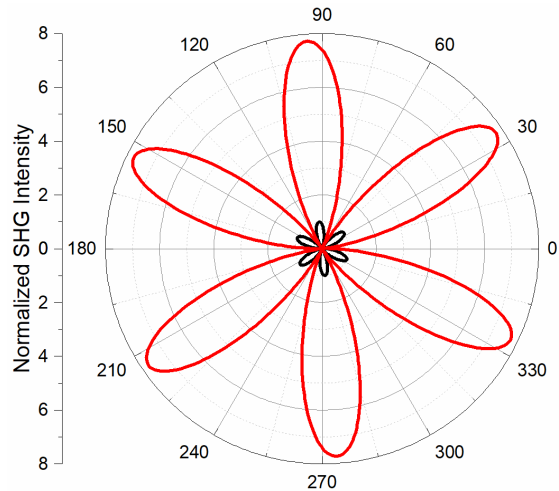

**Figure S10.** Theoretical prediction for MoS2 on FTO (red) and Glass (black) in the perpendicular polarization.

For WS$_2$, theoretically predicting the SHG polarization dependence would only be meaningful in a strain-controlled experiment. Without the nature (uniaxial, biaxial), direction, and local strain amplitude, such calculation would be difficult and of limited applicability. For simplicity, we could assume the monolayer WS$_2$ is under uniaxial strain[10,11], however, a number of free-parameters would have to be implied, and no consistent model or useful data extracted, as more sophisticated methods would be necessary to model the nature of the strain in our samples.

## 8. SHG polarization dependence for TMDCs on Glass

MoS$_2$/Glass and WS$_2$/Glass in the parallel configuration present the same pattern and intensity as in the perpendicular configuration with a phase shift of of 30º, as shown in the Fig. S11. The results for MoS$_2$/Glass can be observed in Fig. S11 (A) and WS$_2$/Glass in Fig. S11 (B).

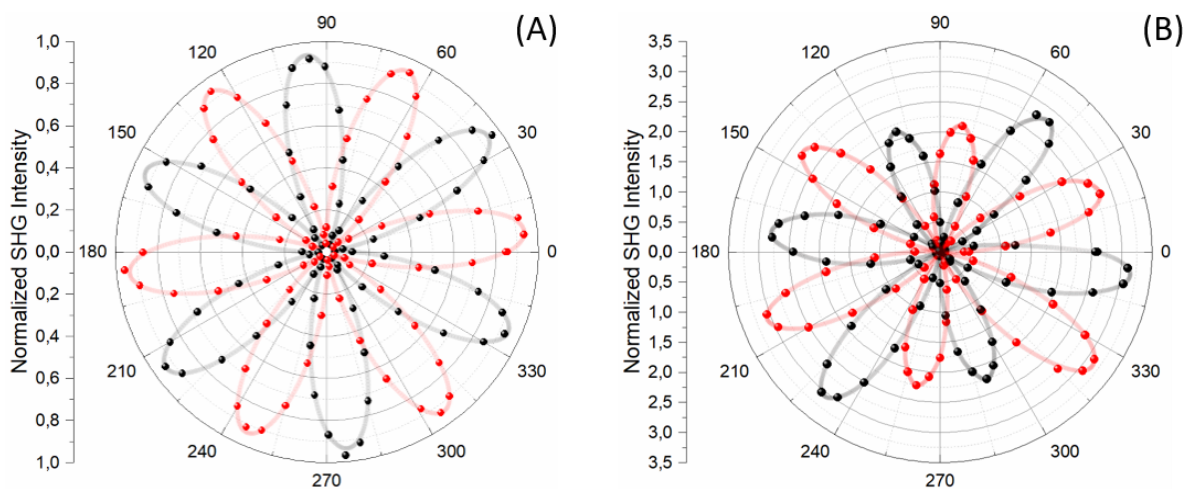

**Figure S11.** Polarized SHG as a function of pump polarization angle in the parallel polarization (red) and perpendicular polarization (black) configurations for MoS$_2$ (A) and WS$_2$ (B).

## 9. Photoluminescence measurements

During sample preparation, TMDC flakes were exfoliated on PDMS, which in contact with the desired substrates and then detached, transferred the monolayers to glass and FTO. The procedure relies on pressure applied during the transfer process, which can possibly deform the viscoelastic stamp, yielding stain[12]. The origin of strain is usually attributed to the inherent lack of stiffness of PDMS. PDMS being soft can get slightly deformed during transfer by the pressure exerted upon contact with the target substrate, likely being the deformation source to induce strain in the flake being transferred[12,13].

In principle, and as already observed by others[14–17], the photoluminescence (PL) shift can be considered an indicator of how the electronic band structure is altered by the application of strain in TMDCs. In our case, PL for monolayer WS$_2$ was measured before and after the transfer process. Figure S12 shows the normalized PL for monolayer WS$_2$ on PDMS and after transferring to FTO (A) and Glass (B). It is possible to observe a shift in the PL peaks, both from 2.02 eV to 2.01 eV, compatible with tensile strain of less than 1% in monolayer WS$_2$[14]. This strain magnitude is compatible with the observed deformation in the SHG polarization dependent plots in Fig. 4(B)[10,11]

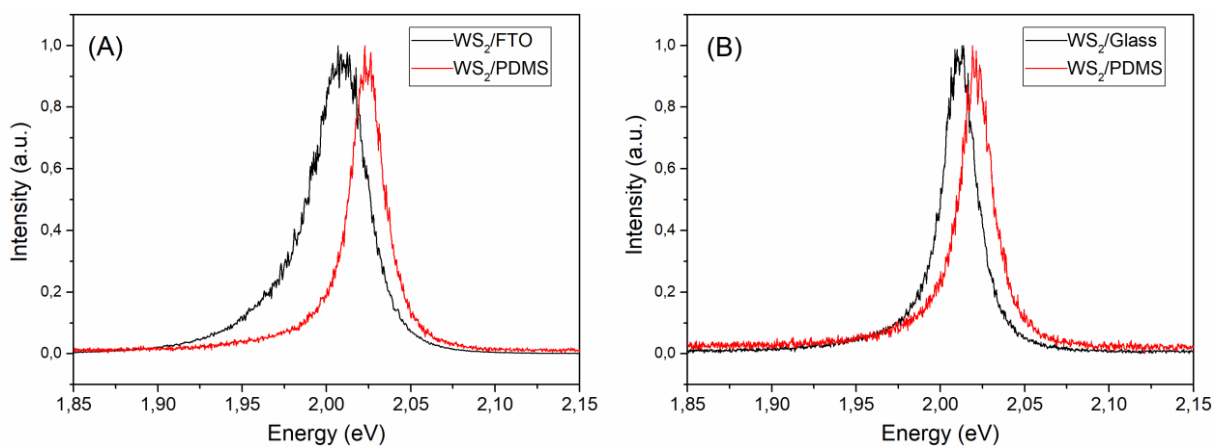

**Figure S12.** Normalized PL for WS$_2$ monolayers on PDMS (red curve) and after transferring to FTO (A) and Glass (B).